\documentclass[ aps, 11pt, final, notitlepage, oneside, nobibnotes, nofootinbib, superscriptaddress, noshowpacs, centertags,twocolumns] {revtex4} \usepackage{graphicx} \usepackage{epstopdf} \usepackage[T2A]{fontenc} \usepackage[english]{babel} \usepackage{natbib}

\def\saoname{Special Astrophysical Observatory,  Russian Academy of Sciences,
              Nizhnii Arkhyz, 369167 Russia}

\def\squareforqed{\hbox{\rlap{$\sqcap$}$\sqcup$}}

\def\sq{\ifmmode\squareforqed\else{\unskip\nobreak\hfil
\penalty50\hskip1em\null\nobreak\hfil\squareforqed
\parfillskip=0pt\finalhyphendemerits=0\endgraf}\fi}

\def\utw{\smash{\rlap{\lower5pt\hbox{$\sim$}}}}

\def\udtw{\smash{\rlap{\lower6pt\hbox{$\approx$}}}}

\def\diameter{{\ifmmode\mathchoice
{\ooalign{\hfil\hbox{$\displaystyle/$}\hfil\crcr
{\hbox{$\displaystyle\mathchar"20D$}}}}
{\ooalign{\hfil\hbox{$\textstyle/$}\hfil\crcr
{\hbox{$\textstyle\mathchar"20D$}}}}
{\ooalign{\hfil\hbox{$\scriptstyle/$}\hfil\crcr
{\hbox{$\scriptstyle\mathchar"20D$}}}}
{\ooalign{\hfil\hbox{$\scriptscriptstyle/$}\hfil\crcr
{\hbox{$\scriptscriptstyle\mathchar"20D$}}}}
\else{\ooalign{\hfil/\hfil\crcr\mathhexbox20D}}%
\fi}}

\newcommand{\aap}{Astron. and Astrophys. }

\newcommand{\aaps}{Astron. and Astrophys. Suppl. }

\newcommand{\aj}{Astron.~J. }

\newcommand{\apss}{Astrophys. and Space Sci. }

\newcommand{\pasp}{Publ. Astron. Soc. Pacific }

\begin{document} \selectlanguage{english} 


%
\title{Possibilities of Using Middle-Band Filters to Search for Polar Candidates} 

\author{\firstname{M.~M.}~\surname{Gabdeev}} \email{gamak@sao.ru} \affiliation{\saoname} 

\author{\firstname{T.~A.}~\surname{Fatkhullin}} \affiliation{\saoname} 

\author{\firstname{N.V.}~\surname{Borisov}} \affiliation{\saoname} 

\begin{abstract} We present a method for searching for polar candidates using mid-band filters. One of the spectral singularities of polars is the $HeII \lambda4686$\AA~ strong emission line. We selected the Edmund Optics filters with central wavelengths of 470, 540, and 656 nm and a transmission bandwidth of 10 nm. These filters cover the regions of the $HeII \lambda4686$\AA~ line, continuum, and the $H_\alpha$ line respectively. We constructed a color diagram based on the available spectra of polars and objects with a zero redshift from the SDSS archive. We show that most polars make a group with unique color indices. In practice, the method is implemented in SAO RAS at the Zeiss-1000 telescope with a new  multi-mode photometer-polarimeter (MMPP). Approbation of the method with the known polars allowed us to develop two criteria to select candidates with an efficiency of up to 75\%. \end{abstract} 

\maketitle

\section{Introduction} Polars are close magnetic cataclysmic systems consisting of a white dwarf whose magnetic field intensity exceeds 10 MG and a red dwarf of a late spectral type. Evolving, the red dwarf fills its Roche lobe and begins to lose matter through the inner Lagrange point L1. If the magnetic field intensity of the white dwarf does not exceed 10 MG, the matter under the gravitational influence of the white dwarf falls on it in Keplerian orbits and forms an accretion disk. In the case when the field intensity exceeds 10 MG, the matter is directed along the magnetic field lines and accrets near the magnetic poles of the white dwarf. In the catalog by Ritter and Kolb\cite{ritt1} there are 148 polars. The number of polars with the determined fundamental parameters is about 20. Such a small number is not enough to prove the existing theoretical concepts of evolution and physics of these systems. Thus, we assigned a task to create a fast and reliable method for searching and classifying  polars. Modern photometry sky surveys, such as the Catalina Sky Survey\cite{drak1}, MASTER\cite{lipu1}, ASSASN \cite{shap1} et al., are aimed at searching for transient and variable objects. To find variability, and much less classification of objects, a large set of observed data is needed. The search for objects by color characteristics is less frequent. Two of them are: the INT Photometric H-Alpha Survey (IPHAS) conducted using the Isaac Newton Telescope\cite{drew1} and the ``Medium-Band Byurakan Survey'' (MBBS)\cite{koto1}. The SDSS sky survey\cite{stou1} in broadband filters allows one to classify the galaxies by redshift. Paula Szkody found the color indices to search for candidate pre-cataclysmic and cataclysmic variables. They are published on the SQL query site according to the SDSS database\footnote{http://skyserver.sdss.org/CasJobs/, the example query ``CVs using colors''}. 

Let us list the observation features of polars in the optical range. In addition to the strong variability of the emission, the classification criterion can be the presence of circular and linear polarization, the $HeII~ \lambda 4686$~ strong emission line comparable in intensity with the $H_\beta$\cite{voih1,patt1} line, and the cyclotron radiation lines. 

We assigned the task to determine the most effective method for searching and classifying the polars with the Zeiss-1000 telescope. The paper describes the possibilities of the method to search for polar candidates using middle-band filters. 

\section{Description and Testing the Method} 
\begin{figure*} \onelinecaptionstrue \captionstyle{normal} \setcaptionmargin{5mm} \includegraphics[width=0.95\columnwidth]{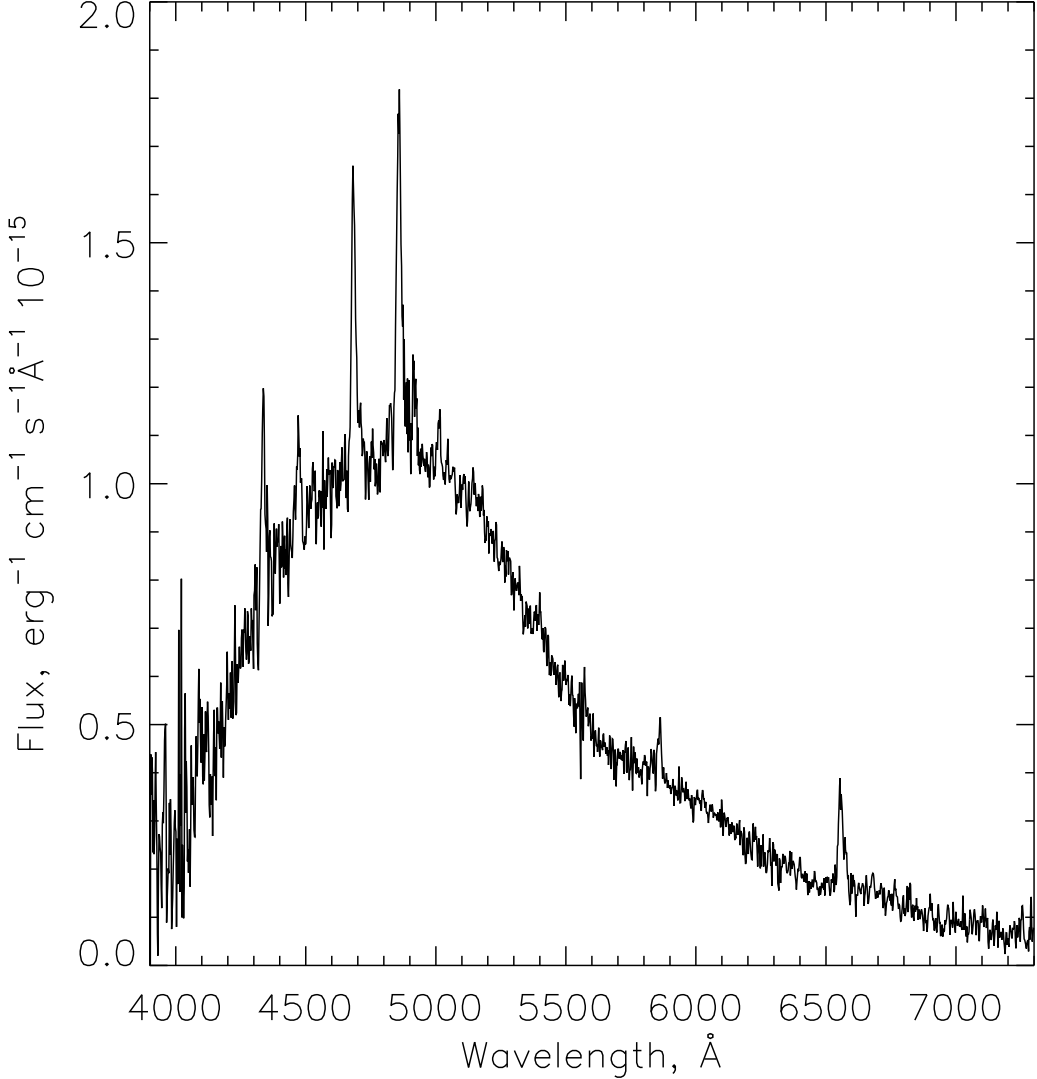} \caption{Spectrum of the polar IPHAS J052832.69+283837.6(published in ASP, but there is no bibtex yet to insert.)\cite{}. A strong cyclotron line is observed in the blue part of the spectrum.} \label{fig_01} \end{figure*}

The middle-band filters of Edmund Optics\footnote{https://www.edmundoptics.com/} were selected for the new Zeiss-1000 photometer. Among them, the filters with central wavelengths of 470 nm and 656 nm. It was previously noted that one of the spectral features of the polars in the optical range is the presence of the $HeII~ \lambda4686$\AA~ strong emission line. The intensity of the ionized helium line is comparable to the intensity of $H_\beta$, the $H_\alpha$ line is more intensive, as a rule. Therefore, comparing the fluxes in two emission lines $HeII~ \lambda4686$\AA~ - filter SED470 and $H_\alpha$ - filter SED656, we can distinguish the polars. To determine the fluxes more correctly, it is necessary to subtract the continuum flux. As was mentioned above, the emission of the polar is strongly variable \cite{crop1,warn1} on the scale of the orbital period and a longer time scale due to a change in the accretion rate. In addition, cyclotron harmonics can be present in the spectra of polars which, in some cases, strongly affect the energy distribution in the optical spectra (Fig. \ref{fig_01}). These factors can affect the effectiveness of the classification of polars. 

\begin{figure*} \onelinecaptionstrue \captionstyle{normal} \setcaptionmargin{5mm} \includegraphics[width=0.95\columnwidth]{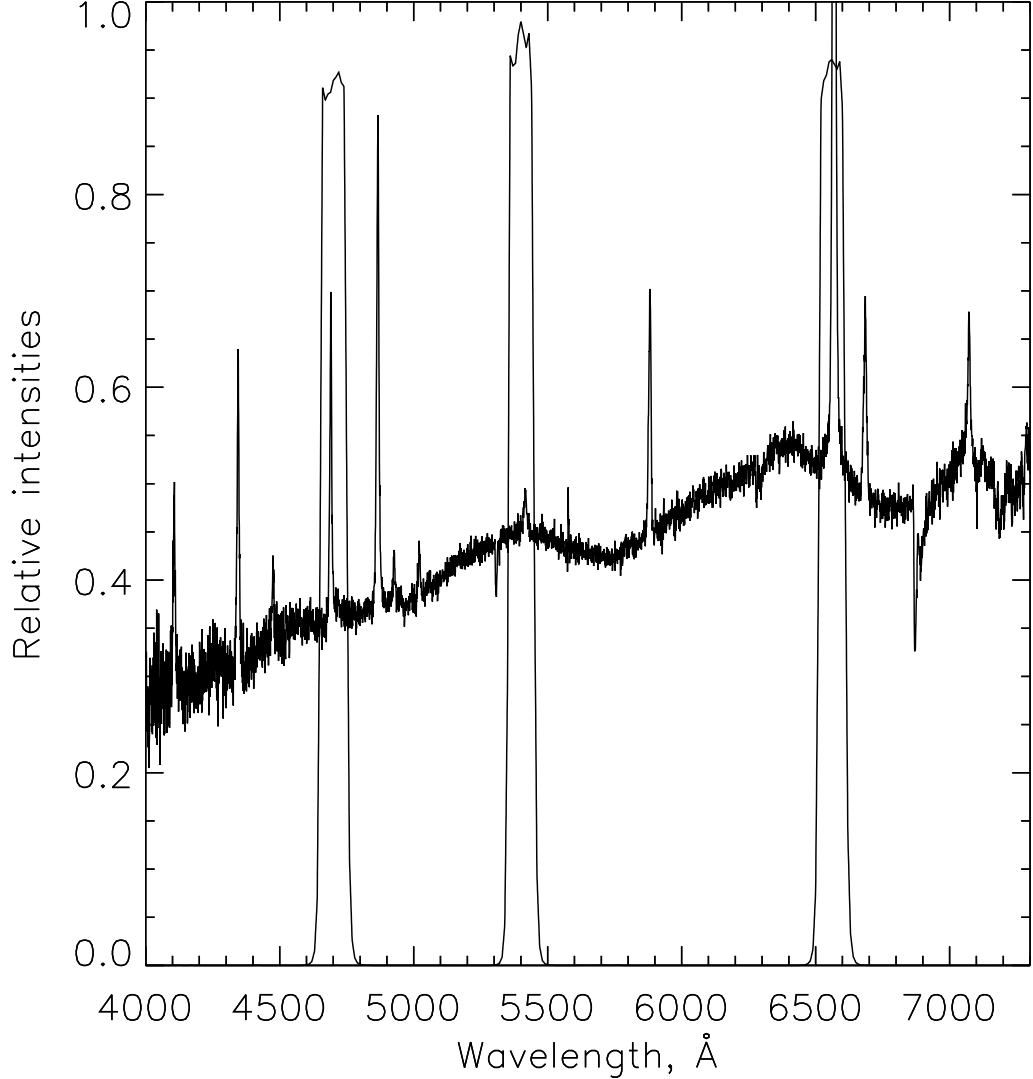} \caption{Spectrum of the polar V808 Aur and transmission curves of the SED470, 540, and 656 filters.} \label{fig_02} \end{figure*}

To test this method, it was necessary to choose a color filter covering the continuum region. In the acquired set of filters, there were the following positions of the central wavelengths (mn): 430, 440, 470, 500, 515, 540, 656, 700. Transmission curves of the color filters in tabular form were provided by Edmund Optics. The bandwidth for all filters is $\Delta\lambda=10$ nm except for SED515 $\Delta\lambda=12.5$ nm. The best division of field stars and polar in the color diagram is achieved using the SED540 filter. Figure \ref{fig_02} shows the transmission bandwidth of the selected filters with the spectra of V808 Aur\cite{bori1} as an example. 

Considering the telescope Zeiss-1000 as a working instrument, we compared the polar classification efficiency using various observation methods. The following parameters were selected to compare the effectiveness of observation methods: the field of view, the total exposure time for a single field, and the limiting magnitude. Table \ref{tab1} shows the parameters listed. Observations with the UAGS spectrograph are not suitable due to the low limiting magnitude. Polarization methods are implemented so that the field of view is only 2'; this greatly reduces the number of objects observed. The average orbital period of the polar is 2 hours, therefore, at least 120 minutes of observation time must be spent on a single field. In addition, it is not always possible to classify a polar according to the light curve. Thus, the method of searching for polar candidates using middle-band filters is the most effective at the Zeiss-1000 telescope. 

\begin{table*}[] \setcaptionmargin{0mm} \onelinecaptionstrue \captionstyle{normal} \caption{Color measurements of the known polars.} \label{tab1} \medskip

\begin{tabular}{|c|c|c|c|c|} \hline

Method &Instrument &Field of view, &Exposure, min& Depth, $^m$\\ \hline

Spectral &UAGS &none &60 &<14\\ \hline

Polarization &MMPP &2 &30& 17-18\\ \hline

Photometric &MMPP &7 &120 &19-20\\ \hline

Color index &MMPP &7 &45 &18-19\\ \hline

\end{tabular} \end{table*} 

To validate the proposed method, a color diagram was constructed to select polars and field stars. We used the polar spectra obtained with the 6-m BTA telescope in the range of $\lambda\lambda=4000-7300$~\AA\AA and spectra of the known polars from the SDSS database. In total, 31 polar spectra were found, 8 from the 6-m BTA telescope archive obtained with the SCORPIO\cite{afan1} and SCORPIO-2\cite{afan2} instruments, 23 from the SDSS archive. The spectra of 1000 objects with a zero redshift were also taken from the SDSS database. This sample can contain not only single stars, but also objects of other types. 

\begin{figure*} \onelinecaptionstrue \captionstyle{normal} \setcaptionmargin{5mm} \includegraphics[width=0.95\columnwidth]{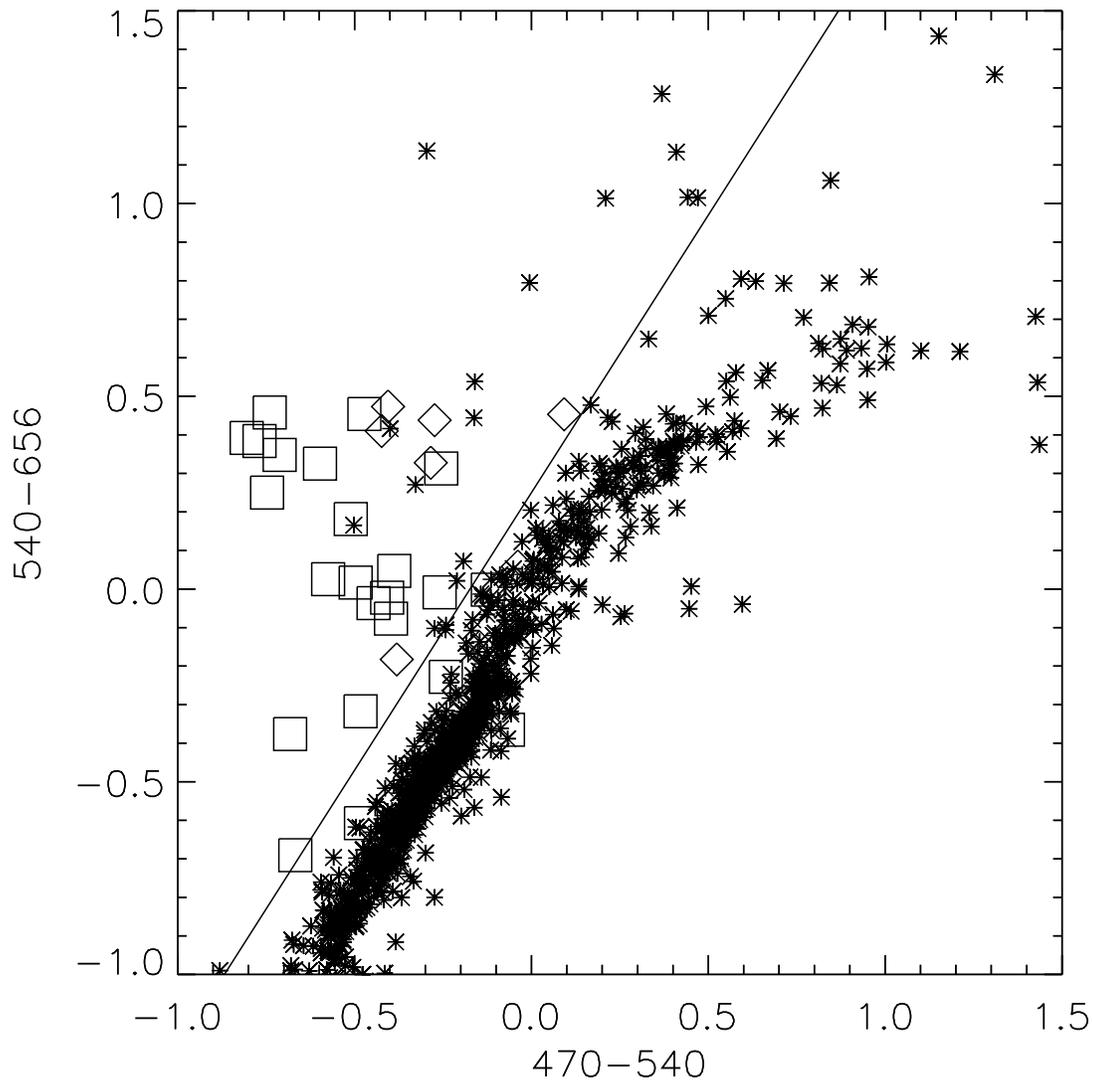} \caption{Color diagram of the sample of polars and field stars. Asterisks denote field stars, diamond symbols indicate the polar spectra obtained with the 6-m BTA telescope, squares indicate the SDSS polar spectra, a line represents the preliminary boundary between polars and field stars. } \label{fig_03} \end{figure*}

The color indices were calculated by folding the spectra with the transmission curves of the SED470, 540, and 656 filters and reducing the fluxes to stellar magnitudes in the AB system. Figure \ref{fig_03} shows the results. Asterisks denote field stars, diamond symbols indicate the polar spectra obtained with the 6-m BTA telescope, squares indicate the SDSS polar spectra, a line represents the preliminary boundary between polars and field stars (SED540-SED656)>1.44*(SED470-SED540)+0.25. The figure demonstrates that the polars, as expected, are separated towards the left part of the diagram. It is worth noting that not all the polars are well distinguished from other objects, this is due to the reasons given earlier in the first paragraph. Preliminarily, the boundaries of the identification region of the polars are drawn. 

\subsection{Observations} Observations were carried out at the new MMPP (Multi-Mode Photometer Polarimeter) of the Zeiss-1000 SAO RAS telescope. During our observations, the photometer was in the test operation mode. A Raptor Eagle V camera with the EEV42-40 chip (2048x2048 pixels) was used as a radiation receiver. The field of view is 7' with a resolution of 0.''333/pixel. The chamber operating temperature, -100 $C^o$, is achieved by water cooling. The observations were carried out in December 2018, on the 2nd and 18th. 

One observation cycle of an object consists of three frames in the SED470, 540, and 656 filters. To eliminate random factors, three cycles are conducted. The spectrophotometric standard, flat fields (FLAT), and electronic zero frame set (BIAS) are also shot every night. The dark current (DARK) is not taken, since its level for the camera is smaller than 1$e^-$ per hour, and the exposure time does not exceed five minutes. 

A sample of the known polars was taken for practical verification of the method. Table \ref{tab2} gives the list of objects and measurements. 

\begin{table*}[] \setcaptionmargin{0mm} \onelinecaptionstrue \captionstyle{normal} \caption{Color measurements of the known polars. ``CCD-failure'' means that a cycle file has been created after exposure, but the information was not recorded, ``no detection'' means that the object could not be detected in one of the filters, ``sunrise'' means that the field was taken at the end of the night, the object was not detected. } \label{tab2} \medskip

\begin{tabular}{|c|c|c|c|c|c|c|} \hline

Object& \multicolumn{2}{|c} {First cycle colours} &\multicolumn{2}{|c} {Second cycle colours} &\multicolumn{2}{|c|} {Third cycle colours} \\ \cline {2-7} name &SED470-SED540 &SED540-SED656 &SED470-SED540 &SED540-SED656 &SED470-SED540 &SED540-SED656 \\ \hline

AN UMa&0.058&0.315&\multicolumn{2}{c|}{CCD-failure}&-0.076&-0.157 \\ \hline

BY Cam&\multicolumn{2}{c|}{CCD-failure} &-0.064&0.892&0.89&1.14\\ \hline

PT Per&-0.04&1.01&0.311&1.00&0.61&1.37 \\ \hline\ GG Leo&0.66 &0.80&-0.37&0.08&0.16&0.69\\ \hline

LW Cam&-0.15&0.74&\multicolumn{4}{c|}{no detection in SED470, SED540}\\ \hline

WX LMi&0.83 &0.87&0.53&0.86&0.55 &0.67\\ \hline

V1309 Ori&0.05&0.42&-0.17&0.75&0.02&0.55 \\ \hline

FR Lyn&0.05 &0.93 &\multicolumn{4}{c|}{sunrise}\\ \hline

V808 Aur&-0.05 &0.03 &0.03&-0.09&\multicolumn{2}{c|}{eclipse}\\ \hline

J0759+1914&0.46 &1.17 &\multicolumn{2}{c|}{CCD-failure}&-0.07&0.85\\ \hline

J0859+0537&-0.02&1.32 &0.10&0.72&\multicolumn{2}{c|}{CCD-failure}\\ \hline

J0733+2619&0.03 &0.85 &\multicolumn{4}{c|}{no detection in SED470, SED540}\\ \hline

J0922+1333&1.13 &0.83 &0.52&1.11&0.48&1.12\\ \hline

J0953+1458&-0.74 &0.72 &-0.20&0.35&0.11&.43 \\ \hline

IPHAS0528&-0.10 &0.35 &0.11&0.23&0.43&0.22\\ \hline

\end{tabular} \end{table*} 

The observations were reduced in the IDL\footnote{https://www.harrisgeospatial.com/Software-Technology/IDL} and Python\footnote{https://www.python.org/} environments using the software for automatic search and photometry of field objects of astronomical images SExtractor\cite{bert}\footnote{https://www.astromatic.net/software/sextractor}. We performed the primary reduction of frames, the deduction of the electronic zero (BIAS), flat-field division (FLAT), and the cosmic ray removal. We will describe the algorithm of the software for data reduction in more detail in a separate paper. 

\subsection{Observation Results} 
\begin{figure*} 
\onelinecaptionstrue 
\includegraphics[width=0.6\columnwidth]{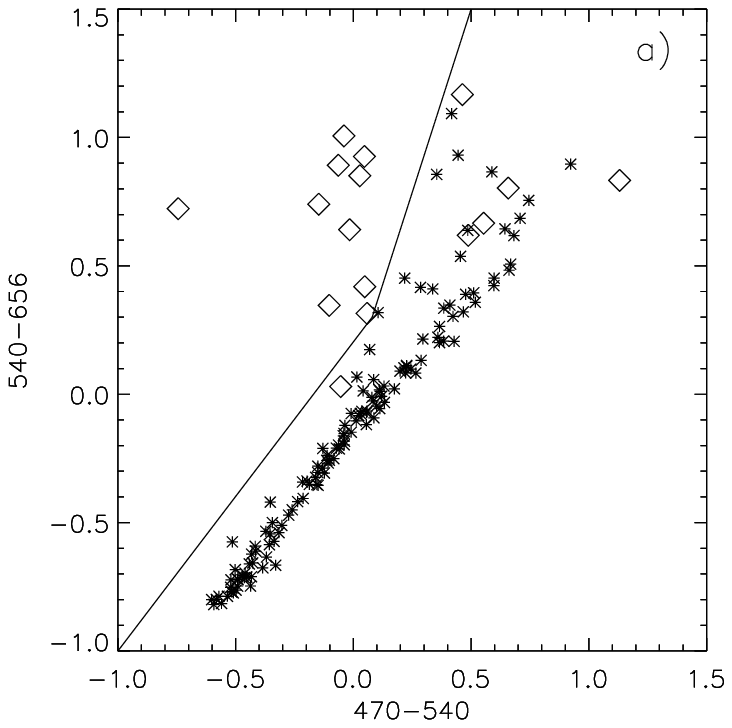} \includegraphics[width=0.6\columnwidth]{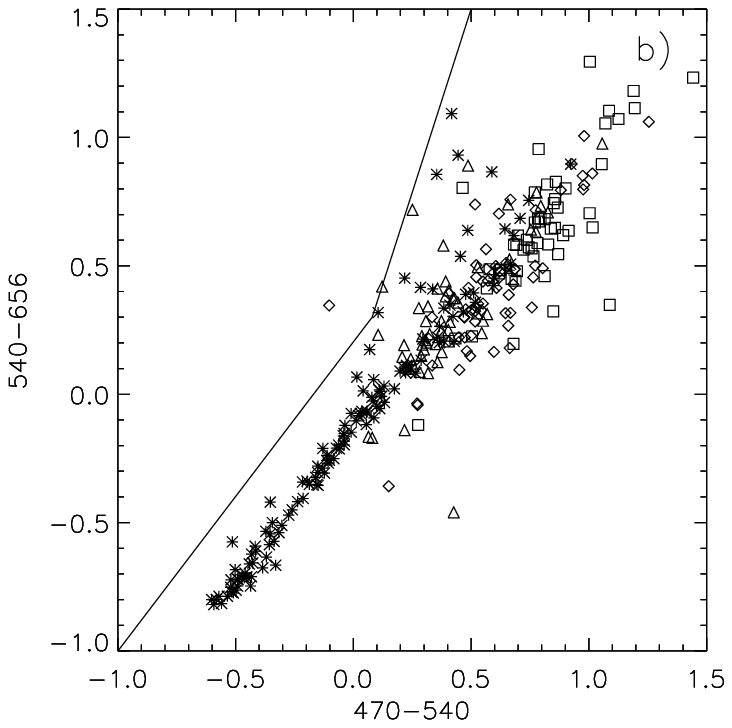} 
\caption{a) Color diagram of stars of different spectral types from the Pickles library (asterisks) and of polars from Table \ref{tab1} (diamond symbols). b) The color diagram of stars of different spectral types from the Pickles library (asterisks) and the objects with the image of the field of the polar IPHAS0528 in different cycles. Cycle 1 -- diamond symbols, cycle 2 -- triangles, cycle 3 -- squares.}\label{fig_04} \end{figure*}

Table \ref{tab1} and Figure \ref{fig_04} show the measurements obtained for the polar sample. The left panel shows the measurements of the polar colors. As a comparison, we give the colors of single stars of different spectral types from the Pickles library\cite{pick1}. The SDSS objects were not reused, as they were selected only by the redshift parameter, and might contain the objects of other types. Panel \ref{fig_04}a shows that the polars are separated from the color sequence of single stars. The average error for the star $19.5^m$ is $0.^m1$. For most field stars, the errors do not exceed the size of characters in the color diagram. It is worth noting that the color characteristics of the polars in the selected filters are variable (see Table \ref{tab1}). The colors of the field stars are concentrated in the red region of the diagram coinciding with the sequence of single stars (Fig. \ref{fig_04}b). Thus, two criteria for the selection of polars can be distinguished: 1) the location relative to the sequence of field stars and 2) the variability of color indices with time. 

Analysis of the results showed that 9 objects from the sample presented are unambiguously distinguished by the first criterion. Three objects, J0922+1333, AN UMa, and V808 Aur, -- by the second criterion (see Table \ref{tab1}). Note that these polars were selected only based on the possibility of observations at a given season. The authors did not know the current spectroscopic characteristics and the accretion state of these objects at the time of observation. 12 of the 16 polars are distinguished by the proposed criteria. Therefore, the proposed method allows us to detect up to 75\% of the polars. The quantitative selection criteria are given in Conclusion. 

\section{Conclusion}
The paper presents a method for searching for polar candidates using the middle-band filters. To implement the method, the Edmund Optics filters with central wavelengths 470, 540, 656 nm and a transmission width of 10 nm were chosen. The first test of the method was carried out by folding the spectra of the known polars obtained with the 6-m BTA telescope and from the SDSS sky survey archive with the transmission curves of the color filters. The comparison was carried out with 1000 objects from the SDSS archive with a zero redshift. It is worth noting that not only single stars could be included in this sample. Most polars are grouped in the upper left part of the diagram. 

In practice, the method was applied at the new MMPP polarimeter of the Zeiss-1000 telescope of SAO RAS. Test observations of the known polars were carried out and a color diagram was constructed. The color characteristics of the polars in the selected filters differ from those of the field stars. In addition, colors of the polars are variable. Two criteria are proposed to classify the polars using this color diagram. These criteria allow one to detect up to 75\% of polars. With accumulation of the observed data, the criteria will be specified. 

\begin{enumerate} \item {Color index $(SED540-SED656) > 1.2\times (SED470-SED540) + 0.2$, when $(SED470-SED540) < 0.08$ and $(SED540-SED656) > 2.86\times (SED470-SED540) + 0.07$, when $(SED470-SED540) > 0.08$} \item {Color variation of an object is three times higher than the average error in measuring color indices.} \end{enumerate} 

In addition to its high detection efficiency, the method has other advantages. Long-term observations are not necessary to create a light curve. The observation time of a single region does not exceed an hour. We intend to observe the selected objects from the CRTS\cite{drak1} sky survey. The CRTS database has thousands of cataclysmic variable candidates. According to the available calculations of the population synthesis of close low-mass binary systems, 10\% of them should be magnetic \cite {lipu1}. Moreover, due to higher spatial resolution of the images from our survey, it is possible to observe regions near the Milky Way. These regions, containing by an order greater number of stars, are ignored by wide-field long-term photometric surveys. 

The implementation of the method will be continued at the new MMPP photometer of the Zeiss-1000 telescope. The observed results are stored in the 3BS\footnote{http://www.sao.ru/3BS/} database. 

\begin{acknowledgments} This work is supported by the Russian Science Foundation (RSF 18-72-00106). \end{acknowledgments}

\end{document}